# Stability of Oxygenated Groups on Pristine and Defective Diamond Surfaces

Eliezer Oliveira[1,2], Chenxi Li[3], Xiang Zhang[3], Anand Puthirath[3], Mahesh R. Neupane[4], James Weil[4], A. Glen Birdwell[4], Tony Ivanov[4], Seoyun Kong[3], Tia Grey[3], Harikishan Kannan[3], Robert Vajtai[3], Douglas Galvao[1,2], Pulickel Ajayan[3]

[1]Gleb Wataghin Institute of Physics, University of Campinas (UNICAMP), Campinas, SP, Brazil.
[2]Center for Computational Engineering & Sciences (CCES), University of Campinas (UNICAMP), Campinas, SP, Brazil.
[3]Department of Materials Science and Nanoengineering, Rice University, Houston, TX, 77005, USA
[4]CCDC US Army Research Laboratory, Adelphi, MD, USA

*The surface functionalization of diamond has been extensively studied through a variety of techniques, such as oxidation. Several oxygen groups have been correspondingly detected on the oxidized diamond, such as C–O–C (ester), C=O (ketonic), and C–OH (hydroxyl). However, the composition and relative concentration of these groups on diamond surfaces can be affected by the type of oxygenation treatment and the diamond surface quality. To investigate the stability of the oxygenated groups at specific diamond surfaces, we evaluated through fully atomistic reactive molecular mechanics (FARMM) simulations, using the ReaxFF force field, the formation energies of C=O, C–O–C, and C-OH groups on pristine and defective diamond surfaces (110), (111), and (311). According to our findings, the C–OH group has the lowest formation energy on a perfect (110) surface, while the C–O–C is favored on a defective surface. As for the (111) surface, the C–O–C group is the most stable for both pristine and defective surfaces. Similarly, C–O–C group is also the most stable one on the defective/perfect (311) surface. In this way, our results suggest that if in a diamond film the (110) surface is the major exposed facet, the most adsorbed oxygen group could be either C–OH or C–O–C, in which the C-O-C would depend on the level of surface defects.*

**INTRODUCTION:**

Recently, the oxidation of polycrystalline diamond film (PCD) was tested using wet and dry methods, such as $H_2SO_4/HNO_3$ acid with various ratios at different temperatures, aqua regia, Fenton solution, Hummer's method, piranha solution, $O_2$ plasma, and UV ozone [1,2]. The best oxidation performance was reached when it was used $H_2SO_4/HNO_3$ in a ratio 5:1 at 360°C, achieving an oxygen content close to 10%.

High-resolution spectrum of XPS shows that various carbon-oxygen groups can be found on PCD diamond surface depending on the oxygenation treatment used, such as C-O-C, C=O, and C-OH. Some dry oxygenation methods, such as $O_2$ plasma-treated produce a high concentration of C-O-C, while UV ozone treatment results in C-OH in majority on the surface [1]. As for the wet oxygenation method, the C-OH group is produced in higher concentration in comparison with the other carbon-oxygen groups.

As can be seen, depending on the different oxidation treatment used, different carbon-oxygen groups can be found in majority on the PCD diamond surface Also, depending on the diamond facets exposed on PCD, as well defects on the surface can influence the preference of a specific carbon-oxygen groups formation. Then, in this work, we performed a theoretical study using fully atomistic reactive molecular mechanics (FARMM) to explore the possible reasons that can affect the carbon-oxygen group formation on PCD surface. Our results suggest that, due to the presence of different diamond surfaces exposed to the oxidation process on PCD, as well the presence of superficial defects, are the dominant reasons to alter the proportion of different oxygenated group formed at the diamond surface.

**MATERIAL AND METHODS**

To explore the energetic stability of C=O, C-O-C, and C-OH oxygenated groups that could be found on the diamond surfaces after the oxygenation process, we have performed fully atomistic reactive molecular mechanics (FARMM) simulations using ReaxFF [3]. Regarding the diamond surfaces that will be used in our study, we selected the surfaces (110), (111), and (311) because they were the most observed in the PCD used in the experiments [1]. Figure 1(a) presents the models for these pristine diamond surfaces. For each diamond surface, we created slabs ~380 carbon atoms and a size of ~14x14x10 Å. The positions of the carbon atoms in the bottom two layers were kept constrained in the diamond bulk positions to mimic bulk-like slabs.

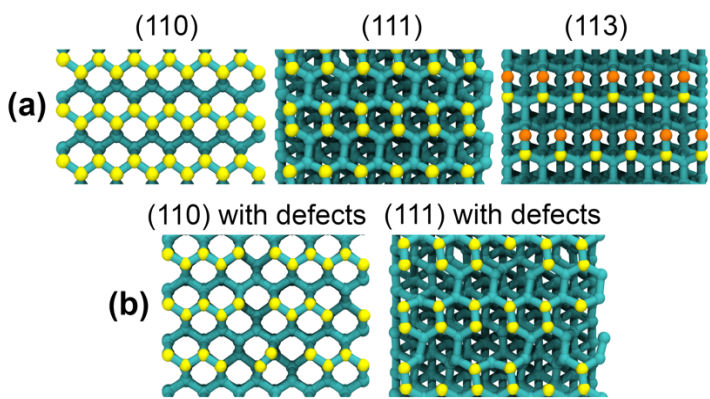

Figure 1. (a) Pristine bare (110), (111), and (311) diamond surfaces in a top and perspective views. (b) Defective (110) and (111) diamond surfaces. In (113) surface, the yellow and orange atoms represent the ones with one and two dangling bonds, respectively. For the other cases, the yellow atoms represent the most exposed carbon atoms.

The geometrical optimizations were performed with Conjugated Gradient (CG) technique, adopting the energy and force convergence tolerances of 0.001 kcal/mol and 0.5 kcal/mol/Å, respectively.

To investigate the energetic stability of each oxygenated group tested in this work, after the geometry optimizations we estimated the binding energy of them using the following equation:

$$BE_{(OG)} = E_{(D+OG)} - E_{(D)} - E_{(OG)} \quad (1)$$

in which the $BE_{(OG)}$ is the binding energy of the oxygenated group, $E_{(D+OG)}$ is the total energy of the diamond slab with the chemisorbed oxygenated group, $E_{(D)}$ is the energy of the bare diamond slab, and $E_{(OG)}$ is the total energy of the oxygenated group. Using the binding energy as defined in Equation 1, the more negative is $BE_{(OG)}$, the more stable will be the specific oxygenated group on the diamond surface. All these simulations were done with the LAMMPS software [4] and the periodic boundary conditions (PBC) were allowed in x and y directions to consider infinite slabs.

## RESULTS AND DISCUSSIONS

As can be seen in Figure 1, a perfect (110) surface does not exhibit any surface reconstruction and the most exposed atoms form a zig-zag configuration, which results in one dangling bond in each exposed carbon atom. To test the stability of the C=O, we can see that it is necessary to have two dangling bonds on the carbon atom. Then, we created a (110) surface with ~20% of random carbon vacancies at the carbon zig-zag configuration in order to have carbon atoms at the surface with two dangling bonds. This defective diamond surface is shown in Figure 1(b), in which the atoms highlighted in red represent the most exposed ones. We tested the C-O-C, C=O, and C-OH one the defective (110) surface in the configurations shown in Figures 2(a)-(e). In the case of the C-O-C and C-OH groups, two different configurations were tested (see Figures 2(a) and (b), and 2(d) and (e)). After the geometry optimizations, we found the following bond-lengths: 1.45 Å for C-O bonds in C-O-C-1 (Figure 2(1)); 1.60 Å for C-O bonds in C-O-C-2 (Figure 2(b)); 1.29 Å for C=O bond (Figure 2(c)); 1.41 and 0.98 Å for, respectively, C-O and O-H bonds in C-OH-1 (Figure 2(d)); and 1.46 and 0.95 Å for, respectively, C-O and O-H bonds in C-OH-2 (Figure 2(e)).

In Table 1 it is presented the binding energies for the these oxygenated groups on the (110) surface. As can be seen, the C-O-C-1 (Figure 2(a)) is the most stable one, followed by the C=O, C-OH-1 (Figure 2(d)), C-O-C-2 (Figure 2(b)), and C-OH-2 (Figure 2(e)). We see that the presence of defects on the (110) surface change the stability of the oxygenated groups. Considering a perfect (110) surface, in which is possible to have only C-O-C and C-OH groups, the C-OH will be the most stable one.

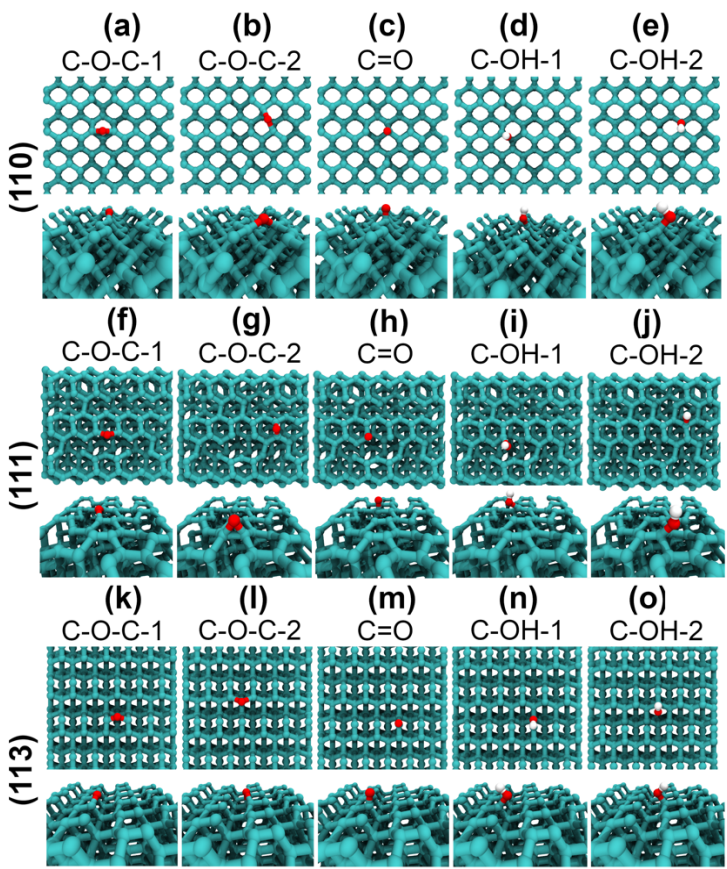

Figure 2. Configurations for the carbon-oxygen groups on diamond surfaces used for binding energy estimates: (a) C-O-C in conformation 1 (C-O-C-1), (b) C-O-C in conformation 2 (C-O-C-2), (c) C=O, (d) C-OH in conformation 1 (C-OH-1), and (e) C-OH in conformation 2 (C-OH-2) on a (110) surface; (f) C-O-C in conformation 1 (C-O-C-1), (g) C-O-C in conformation 2 (C-O-C-2), (h) C=O, (i) C-OH in conformation 1 (C-OH-1), and (j) C-OH in conformation 2 (C-OH-2) on a (111) surface; (k) C-O-C in conformation 1 (C-O-C-1), (l) C-O-C in conformation 2 (C-O-C-2), (m) C=O, (n) C-OH in conformation 1 (C-OH-1), and (o) C-OH in conformation 2 (C-OH-2) on a (311) diamond surface. The cyan, red, and white atoms are, respectively, carbon, oxygen, and hydrogen.

Considering the surface (111), its most stable configuration presents a surface reconstruction 2x1 [5–9], which results in carbon atoms with one dangling bond each at the surface. As discussed before, to form a C=O bond, it is necessary a carbon atom with 2 dangling bonds. To circumvent this issue, we created a (111) surface with with ~20% of random carbon vacancies, which results in superficial carbon atoms with up to 2 dangling bonds (see Figure 1(b), in which the most exposed carbon atoms on the surface are highlighted in yellow). In Figures 2(f) to (j) it is presented the tested configurations for the C-O-C, C=O, and C-OH groups at the (111) surface. Here we also tested two different configurations for the C-O-C and C-OH groups (see Figures 2(f) and (g), and Figures 2(i)-(j)). The bond-length values obtained in these cases were: 1.50 Å for C-O bonds in C-O-C-1 in Figure 2(f); 1.44 Å for C-O bonds in C-O-C-2 in Figure 2(g); 1.41 Å for C=O bond shown in Figure 2(h); 1.41 and 0.98 Å for C-O and O-H bonds, respectively, in C-OH-1 in Figure 2(i); and 1.48 and 0.93 Å for C-O and O-H bonds, respectively, in C-OH-2 in Figure 2(j). From the estimated binding energies presented in Table 1, we can see that the C-OH-1 (Figure 2(i)) is the most stable on surface (111), followed by the C-O-C-2 (Figure 2(g)). The C-O-C-1, C=O, and C-OH-2 are not stable on this surface.

On the surface (311), we see that it is present carbon atoms with one and two dangling bonds (see Figure 1(a) in which the carbon atoms with one dangling bond are highlighted in yellow, and with two dangling bonds are highlighted in orange). In Figures 2(k)-(o) we present the C-O-C, C=O, and C-OH on the (311) surface in the configurations shown in Figures 4(k)-(o). From the energy minimized structures, we can find the following bond-lengths: 1.44 Å for C-O bonds in C-O-C-1 in Figure 2(k); 1.60 Å for C-O bonds in C-O-C-2 in Figure 2(l); 1.28 Å for C=O bond in Figure 2(m); 1.40 and 0.98 Å for, respectively, C-O and O-H bonds in C-OH-1 in Figure 2(n); and 1.45 and 0.96 Å for, respectively, C-O and O-H bonds in C-OH-2 in Figure 2(o). According to the binding energies presented in Table 1, the most stable group is the C-O-C-1 (Figure 2(k)), followed by the C=O (Figure 2(m)) and C-OH-2 (Figure 2(o)). The C-O-C-2 (Figure 2(l)) and C-OH-1 (Figure 2(n)) are not stable in surface (311).

Table 1. Theoretical binging energies of the C-O-C, C=O, and C-OH groups on (110), (111), and (311) diamond surfaces. See in Figure 2 the configurations of each carbon-oxygen group at the diamond surface.

| Surface | Binding energy (eV) | | | | |
|---|---|---|---|---|---|
| (110) | C-O-C-1 | C-O-C-2 | C=O | C-OH-1 | C-OH-2 |
|  | -8.49 | -3.93 | -6.58 | -4.05 | -3.33 |
| (111) | C-O-C-1 | C-O-C-2 | C=O | C-OH-1 | C-OH-2 |
|  | 3.32 | -4.93 | 14.06 | -5.40 | 7.23 |
| (311) | C-O-C-1 | C-O-C-2 | C=O | C-OH-1 | C-OH-2 |
|  | -9.22 | 2.78 | -6.63 | 1.49 | -1.07 |

# CONCLUSIONS

According to our study, the C-OH group is the most stable one when we are considering a (110) and (111) perfect (non-defective) diamond surfaces. However, the C-OH groups may not be the most stable ones when is it taken in to account defects on the surface. In the case of the (311) surface, we see that the C-O-C group is the most stable, but in general this surface is less exposed than the (110), which suggests that this surface may not contribute significantly with the accounting of the oxygenated groups in diamond. In general, our results suggests that if the oxidation process cause significant structural damages on the diamond surfaces or if the diamond surfaces have poor quality, other groups other than the C-OH one may be favored to be found in larger quantities.

# ACKNOWLEDGMENTS

The authors thank the Brazilian agencies CNPq and FAPESP (Grants 2013/08293-7, 2016/18499-0, and 2019/07157-9) for financial support. Computational support from the Center for Computational Engineering and Sciences at Unicamp through the FAPESP/CEPID Grant No. 2013/08293-7 and the Center for Scientific Computing (NCC/GridUNESP) of São Paulo State University (UNESP) is also acknowledged.

**References:**

1. C. Li, X. Zhang, E. F. Oliveira, et al., *Carbon* **182**, 4977 (2021).
2. A. B. Putirath, E. F. Oliveira, G. Gao, *et al.*, *Chem. Mater.* **33**, 4977 (2021).
3. A. C. T. van Duin, S. Dasgupta, F. Lorant, W. A. Goddard, *J. Phys. Chem. A* **105**, 9396 (2001).
4. S. J. Plimpton, *Comput. Phys.* **117**, 1 (1995).
5. S. J. Sque, R. Jones, P. R. Briddon, *Phys. Status Solidi Appl. Mater. Sci.* **11**, 202 (2005).
6. P. Rivero, W. Shelton, V. Meunier, *Carbon* **110**, 469 (2016).
7. M. De La Pierre, M. Bruno, C. Manfredotti, *et al.*, *Mol. Phys.* **112**, 1030 (2014).
8. H. Kawarada, *Surf. Sci. Rep.* **26**, 205 (1996).
9. E. F. Oliveira, M. R. Neupane, C. Li, *et al.*, *Comput. Mater. Sci*. **200**, 110859 (2021).